\documentclass[pss,fleqn]{w-art}
\usepackage{times}
\usepackage{w-thm}
\usepackage[]{graphicx}
\begin{document}
\DOIsuffix{theDOIsuffix}
\Volume{XX}
\Issue{1}
\Copyrightissue{01}
\Month{01}
\Year{2004}
\pagespan{1}{}



\title[Electron-phonon interaction in thin copper and gold films]{Electron-phonon interaction in thin copper and gold films}


\author[J.T. Karvonen]{J.T. Karvonen\footnote{Corresponding
     author: e-mail: {\sf jenni.karvonen@phys.jyu.fi}, Phone: +358 14 260 2442,
     Fax:+358 14 260 2351}\inst{1}} \address[\inst{1}]{NanoScience Center, Department of Physics, P.O. Box 35, 
FIN-40014, University of Jyv\"askyl\"a, Finland}

\author[L.J. Taskinen]{L.J. Taskinen\inst{1}}
\author[I.J. Maasilta]{I.J. Maasilta\inst{1}}
\begin{abstract}
We have studied the electron-phonon (e-p) interaction in thin Cu and Au films at sub-Kelvin temperatures with the help of the hot electron effect, 
using  symmetric normal metal-insulator-superconductor tunnel junction pairs as thermometers. By Joule heating the electron gas and measuring the 
electron and the lattice temperatures simultaneously, we show that the electron-phonon scattering rate follows a $T^{4}$ temperature dependence in 
both metals. The result is in accordance with the theory of e-p scattering in disordered films with vibrating boudaries and impurities, in contrast to 
the $T^{3}$-law expected for pure samples,
and $T^{2}$-law for static disorder.
\end{abstract}      

\maketitle

\section{Introduction}
\sloppy
Interaction between conduction electrons and thermal phonons is elementary for many processes and phenomena at low temperatures, but there are still 
only few experimental studies supporting the theory for typical disordered metals. Several earlier results indicated that both in ordered and 
disordered films the electron-phonon scattering rate follows power law $1/\tau _{e-p} \propto T^{3}$ \cite{Gantmakher,Roukes,well,kansk}, and thus the 
heating power from electrons to phonons is of the form $P=\Sigma\Omega(T_{\rm e}^{5}-T_{\rm p}^{5})$, where $\Sigma$ is a material  dependent 
parameter, $\Omega$ the volume of the sample, $T_{\rm e}$ the electron temperature and $T_{\rm p}$ the phonon temperature. However, according to the 
theory this result is possible only for pure samples in the limit $ql>>1$, where $q$ is the wavevector of the dominant thermal phonons and $l$ the 
electron mean free path, coupling to longitudinal phonons only. 

In disordered metals, where $ql<<1$, theory predicts that the scattering rate from vibrating disorder is $1/\tau _{\rm e-p}\propto T^{4}$ 
\cite{Schmid} so that the heating power is     
\begin{equation}
\label{disorder}
P=\Sigma'\Omega(T_{\rm e}^{6}-T_{\rm p}^{6}),
\end{equation}
when electrons are expected to couple dominantly to transverse acoustical phonons \cite{Sergeev}. This result has not been widely confirmed, and in 
fact we are not aware of any observation of it in standard normal metal films like Cu, Au, Ag, etc. Some  evidence exists for strongly disordered ($l$ 
$\sim$ 1nm) Ti and Hf films \cite{Gerhenson} and Si SOI wafers \cite{kivi}.

Recently we have observed the theoretically expected $T^{4}$ dependence for e-p scattering rate in evaporated copper \cite{Maasilta} and gold films. 
We have measured the rate at which electrons in a normal metal wire overheat, when DC power is applied to it. Overheating rate was determined directly 
by measuring the electron and phonon temperatures with help of SINIS-thermometers.
\section{Experimental techniques and samples}
\sloppy
We have measured several Cu samples with a range of film thicknesses and thus a range of $l$, as reported earlier \cite{Maasilta}. Here we concentrate 
on comparing the data of one of those Cu samples with new data from a Au sample. To assess the quality of metal films and to determine the electron 
mean free path $l$, we measured the sample resistivities $\rho$ at 4.2 K and at room temperature in a four-probe configuration, and  determined $l$ 
for the sub-Kelvin range from the 4.2 K data. Now we can estimate the critical phonon temperatures $T_{\rm p}^{ql=1}$ corresponding to $ql=1$ by 
$T_{\rm p}^{ql=1}=(\hbar c_{t})/(2.82k_{\rm B}l)$, where $c_{\rm t}$ is the transverse sound velocity ($c_{\rm t}^{\rm Cu}$=2300 m/s and $c_{\rm 
t}^{\rm Au}$= 1200 m/s). 
\begin{table}
\centering
\begin{tabular}{|l|l|l|l|l|l|}
\cline{1-6}
\vbox to1,88ex{\vspace{1pt}\vfil\hbox to7,80ex{\hfil sample\hfil}} & 
\vbox to1,88ex{\vspace{1pt}\vfil\hbox to7,80ex{\hfil $t$ (nm)\hfil}} & 
\vbox to1,88ex{\vspace{1pt}\vfil\hbox to7,80ex{\hfil RRR\hfil}} & 
\vbox to1,88ex{\vspace{1pt}\vfil\hbox to7,80ex{\hfil $\rho$ ($\mu\Omega$cm)\hfil}\vfil} & 
\vbox to1,88ex{\vspace{1pt}\vfil\hbox to7,80ex{\hfil $l$ (nm)\hfil}\vfil} & 
\vbox to1,88ex{\vspace{1pt}\vfil\hbox to7,80ex{\hfil $T_{\rm p}^{ql=1}$ (K)\hfil}\vfil} \\

\cline{1-6}
\vbox to1,88ex{\vspace{1pt}\vfil\hbox to7,80ex{\hfil Cu\hfil}} & 
\vbox to1,88ex{\vspace{1pt}\vfil\hbox to7,80ex{\hfil 35\hfil}} & 
\vbox to1,88ex{\vspace{1pt}\vfil\hbox to7,80ex{\hfil 1.8\hfil}} & 
\vbox to1,88ex{\vspace{1pt}\vfil\hbox to7,80ex{\hfil 3.06\hfil}\vfil} & 
\vbox to1,88ex{\vspace{1pt}\vfil\hbox to7,80ex{\hfil 21.4\hfil}\vfil} & 
\vbox to1,88ex{\vspace{1pt}\vfil\hbox to7,80ex{\hfil 0.29\hfil}\vfil} \\

\cline{1-6}
\vbox to1,88ex{\vspace{1pt}\vfil\hbox to7,80ex{\hfil Au\hfil}} & 
\vbox to1,88ex{\vspace{1pt}\vfil\hbox to7,80ex{\hfil 40\hfil}} & 
\vbox to1,88ex{\vspace{1pt}\vfil\hbox to7,80ex{\hfil 2.3\hfil}} & 
\vbox to1,88ex{\vspace{1pt}\vfil\hbox to7,80ex{\hfil 1.94\hfil}\vfil} & 
\vbox to1,88ex{\vspace{1pt}\vfil\hbox to7,80ex{\hfil 43.0\hfil}\vfil} & 
\vbox to1,88ex{\vspace{1pt}\vfil\hbox to7,80ex{\hfil 0.08\hfil}\vfil} \\

\cline{1-6}
\end{tabular}
\caption{Parameters of measured Cu and Au samples. $\rho$ is the value measured at 4.2 K, $T_{\rm p}^{ql=1}$ is for transverse modes.}
\end{table}

\begin{figure}[htb]
\centering
\includegraphics[width=10cm, height=2.3cm]{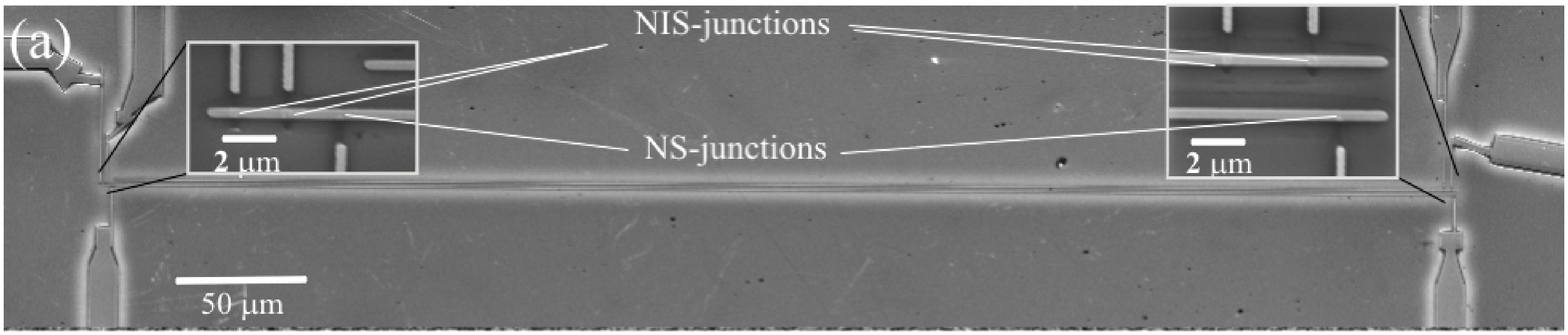}\includegraphics[width=5cm, height=2.3cm]{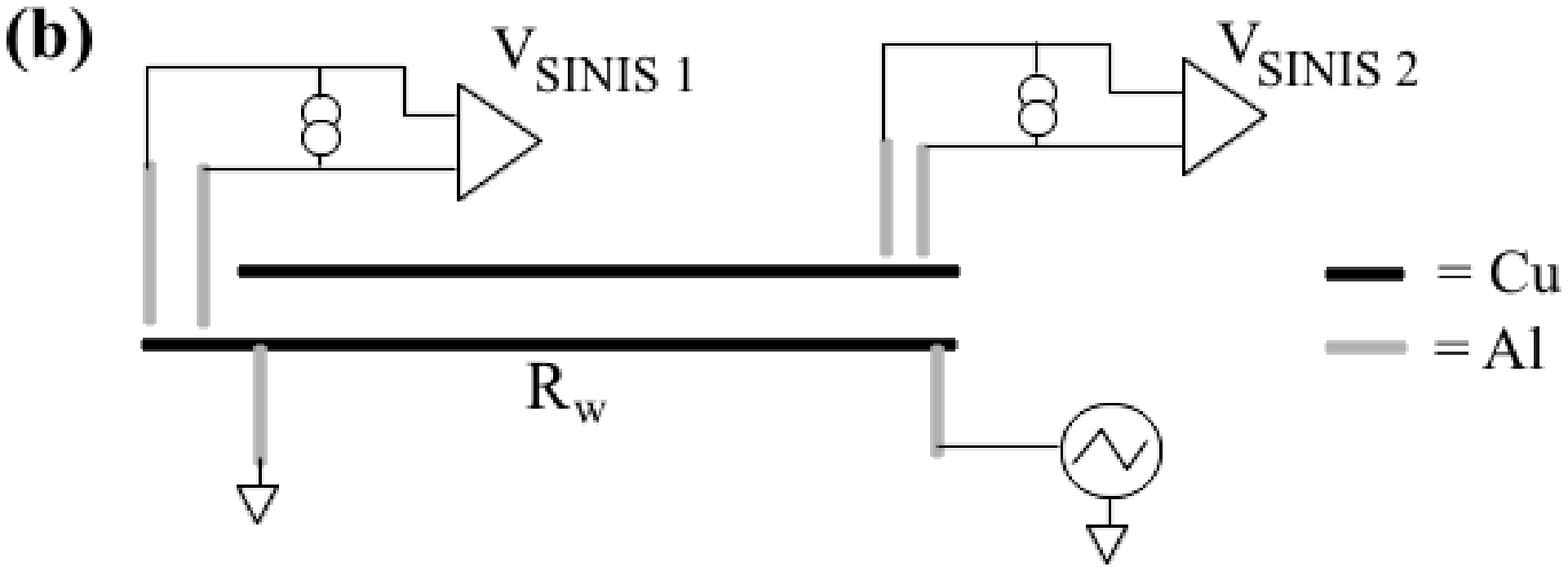}
\caption{(a) An SEM image of a sample. (b) Schematic of the sample and the measurement circuit.}
\label{fig:1}
\end{figure}

All measured samples have two normal metal wires of length $\sim$ 500 $\mu$m and width $\sim$ 300 nm on thermally oxidized Si substrate, separated by 
a distance 2 $\mu$m. The normal metal wires were connected to the measurement circuit by superconducting Al leads as shown in the SEM image 
\ref{fig:1} (a). Two pairs of current biased leads in both wires are SINIS tunnel junctions for measuring the electron temperature of the wire. The 
junctions connecting the lower wire to a voltage source are NS-junctions for heating the normal metal. For low enough heating voltages the 
NS-junctions are biased within the superconducting gap $\Delta$ of the leads, and Andreev reflection can take place. This makes the NS-junctions 
perfect thermal insulators despite being electrically conducting, and thus it is possible to Joule heat the normal metal uniformly through the 
NS-junctions.

All the measurements were performed by current biasing the two SINIS thermometers and measuring their DC voltages simultaneously in a dilution 
refrigerator with a base temperature 60 mK. The thermometers were calibrated by varying the bath temperature of the refrigerator very slowly, while 
measuring  the temperature of the sample stage with a calibrated Ruthenium Oxide thermometer.  
\begin{figure}[tb]
\centering
\includegraphics[width=6.7cm, height=5cm]{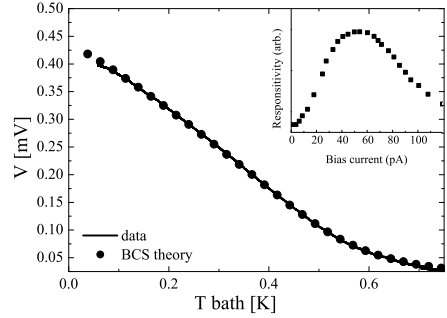}
\caption{Calibration for the SINIS junction in Au sample connected to the lower wire in Fig.\ref{fig:1}. The junction is biased at $I=60$ pA according 
to responsivity $dV/dP$ measurement shown in the inset.}
\label{fig:2}
\end{figure}

Figure \ref{fig:2} shows an example of a calibration measurement. It is evident that the agreement between measured calibration curve and 
corresponding theoretical curve calculated numerically from the BCS theory is good except at the very lowest temperatures. This deviation could arise 
by two major mechanisms:(i) a component of leakage or noise current affects the measurement, and (ii) the electrons overheat at the low-$T$ range. We 
have ruled out (i) by directly measuring the responsitivities $dV/dP$ of the SINIS-thermometers as a function of $I$ using a lock-in amplifier (fig. 
\ref{fig:2} inset), and setting the bias current such that we see the expected behavior from BCS theory ($dV/dP$ decreasing vs. $I$). Mechanism (ii), 
overheating by noise power $P_{\rm noise}$, is an obvious explanation since we have seen that filtering the lines clearly improves the situation. 
Overheating can be modelled with the help of theory for e-p scattering and the Kapitza resistance \cite{Swartz} and the model fits the data perfectly 
\cite{Maasilta}. Thus, the thermometers are calibrated correctly by reading $T_{\rm e}$ from the BCS-theory curve.  
\section{Results}
\sloppy
In our experiment we applied slowly ramping DC voltage to Joule heating the lower wire through NS-junctions and determined heating power by measuring 
the current and voltage directly in a 4-wire configuration at $T_{\rm bath}=60$ mK. Simultaneously we measured the electron temperatures in a both 
wires by SINIS-thermometers. 

\begin{figure}[htb]
\centering
\includegraphics[width=5.3cm, height=4.1cm]{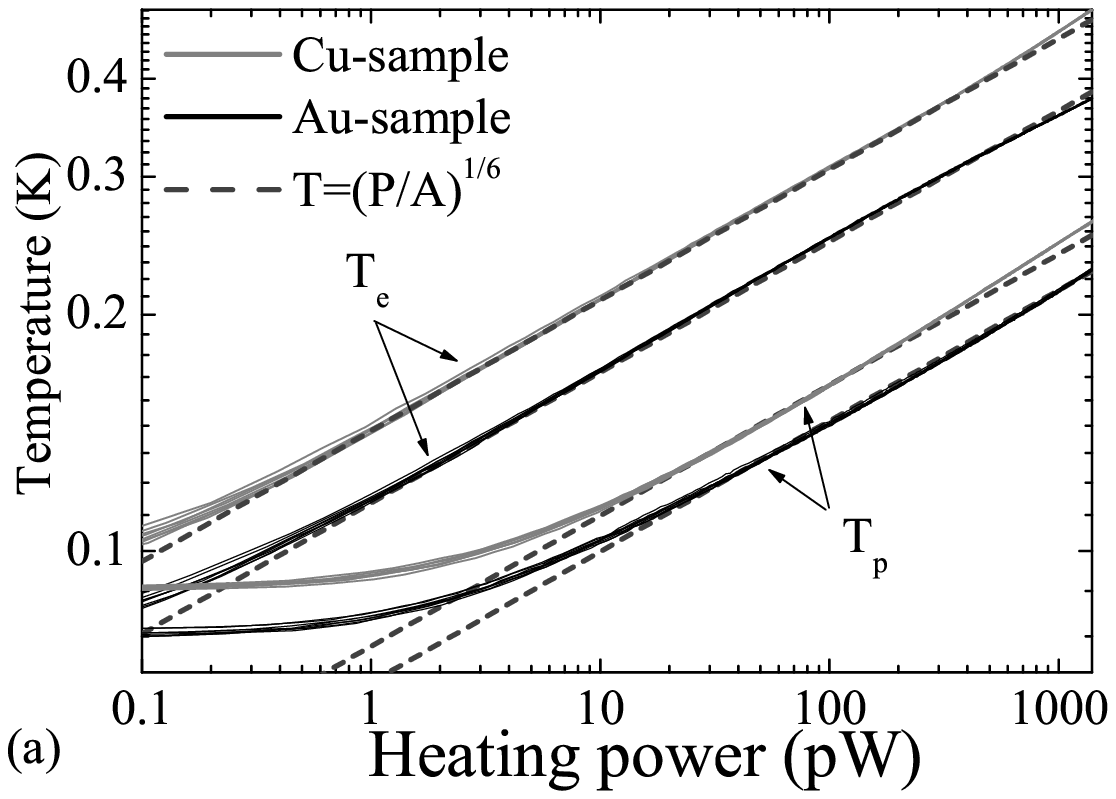}\includegraphics[width=5.3cm, height=4.1cm]{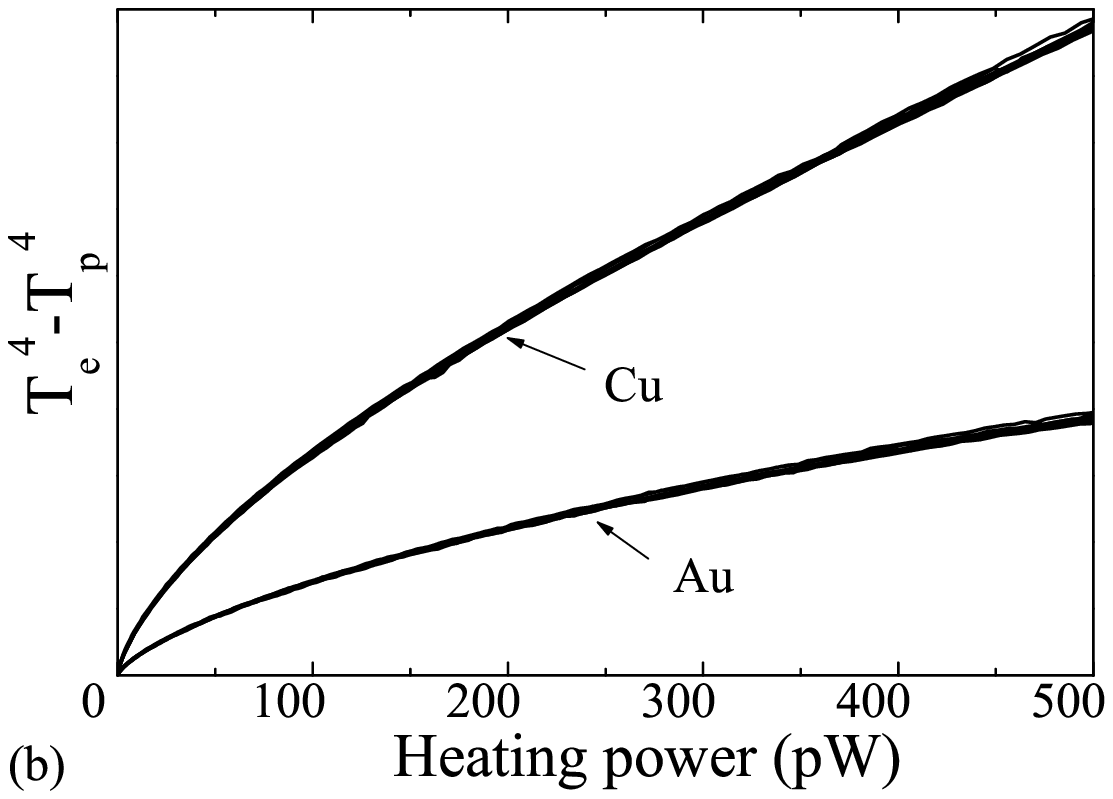}\\
\includegraphics[width=5.3cm, height=4.1cm]{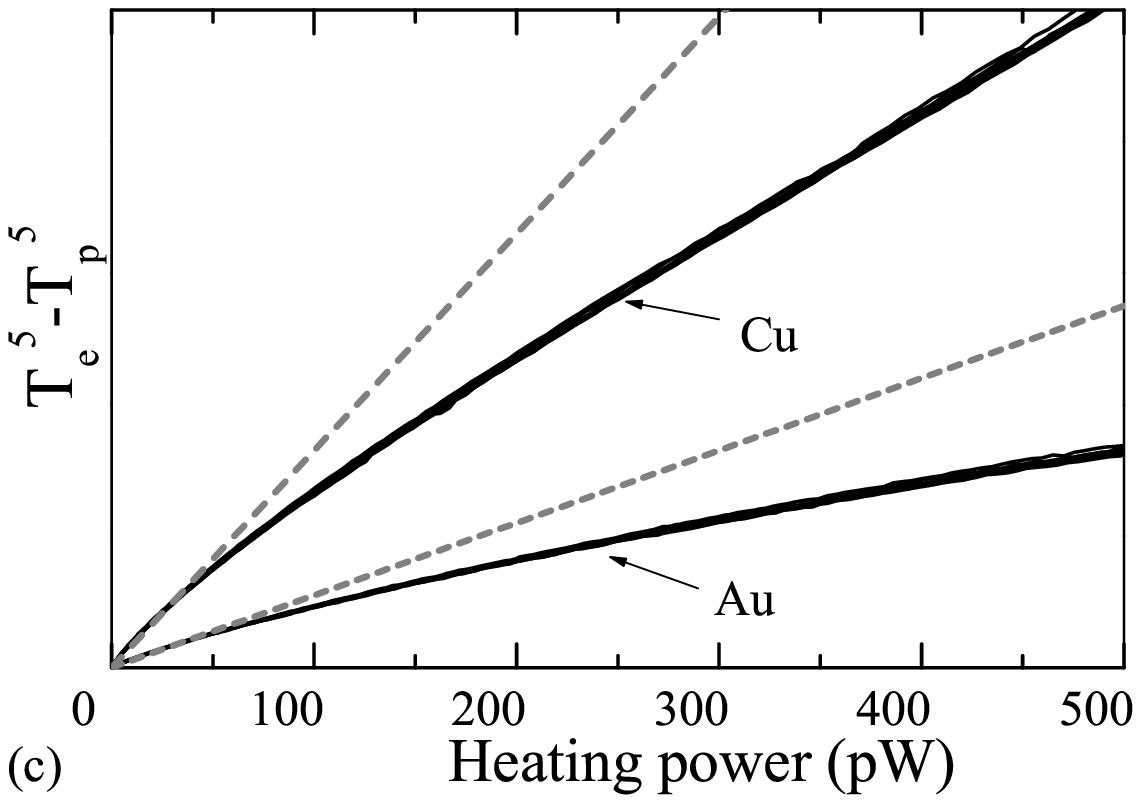}\includegraphics[width=5.3cm, height=4.1cm]{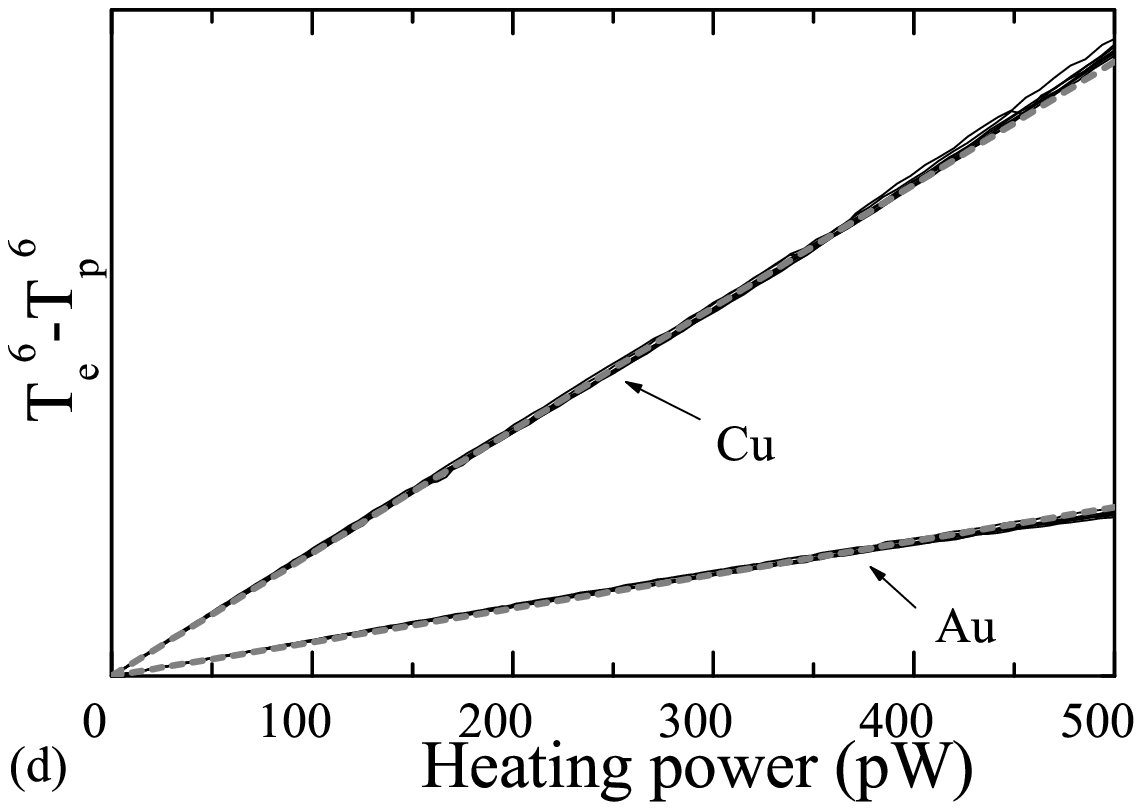}
\caption{(a) The temperature $T_{\rm e}$ and $T_{\rm p}$ vs heating power $P$ in log-log scale. Dashed lines are $T=(P/A)^{1/6}$. In the Figs. (b), 
(c) and (d) $T_{\rm e}^{n}-T_{\rm p}^{n}$ vs $P$ with $n=4,5,6$ is plotted. Dashed lines are one parameter fits through the origin.}
\label{fig:3}
\end{figure}

In Fig. \ref{fig:3} (a) we plot the electron temperatures in both wires vs $P$ obtained from several sweeps for both samples. As can be seen, the 
temperatures of both wires rise when heating power is increased. This means that the second wire is heated indirectly by phonons, since there is no 
direct electrical contact between the wires. Therefore,  there must be mechanism in the substrate to backscatter hot phonons generated in the lower 
wire. Most likely this scattering take place in the oxide layer and/or at the interface between the oxide and the substrate. Looking at Fig. 
\ref{fig:3}(a), above a few pW this phonon heating clearly dominates the signal over the noise heating in the upper, non-Joule heated wire (labelled 
$T_p$). Thus we can identify this as the phonon temperature $T_{\rm p}$, since all the power absorbed from the phonons has to be re-emitted back into 
the substrate. Moreover, since the two wires are separated only by $\sim$ 2 $\mu$m, much shorter than lateral mean free path of phonons in SiO 
\cite{Swartz}, the same $T_{\rm p}$ is seen by the lower wire also. The conclusion is that we can simultaneously measure the electron and phonon 
temperatures of the heated wire.   

Looking back at Fig. \ref{fig:3}(a) we can see that for both samples $T_{\rm e}$ and $T_{\rm p}$ scale as $(P/A)^{1/6}$. Since the approximation 
$T_{\rm e}^{6}>>T_{\rm p}^{6}$ holds for both samples, we can conclude from the behaviour of $T_{\rm e}$ alone  that equation (\ref{disorder}) is 
valid, and e-p scattering is mediated by vibrating disorder. However, empirically the relation $T_{e}^{6}/T_{p}^{6}=const$ also holds for both 
samples, which explains why  in the data of Fig. \ref{fig:3}(a) $P\propto T_{\rm p}^{6}$ also. The physical reason for this empirical observation is 
not clear. 

In figures \ref{fig:3} (b-d) we also plot $T_{\rm e}^{n}-T_{\rm p}^{n}$ vs applied Joule heating power $P$ when $n=4,5,6$ in linear scale, which is 
strictly the only correct way to plot the data without approximations.  It can be clearly seen from Fig. \ref{fig:3} (b) that $n=4$ does not fit the 
data. This means that Kapitza resistance or e-p scattering by static impurities is not in a dominant role. Also the expected theory for pure samples, 
$n=5$, does not fit the data according to the curve shown in Fig. \ref{fig:3}(c). However from Fig. \ref{fig:3}(d) we can see that $n=6$ gives 
straight line for both samples and thus evidence for the validity of equation (\ref{disorder}). From the linear fits through the origin we get 
$\Sigma'=2.4\times 10^{10}$W/K$^6$m$^3$ for the Copper sample and $\Sigma'=7.5\times 10^{10}$W/K$^6$m$^3$ for the Gold sample. It should be stressed 
that in the disordered regime $\Sigma'$ is not only a material dependent parameter but depends directly on the level of disorder through $l$ 
\cite{Sergeev}. For comparisons between different measurements, therefore, both $\Sigma'$ and $l$ need to be quoted. Also, it is unclear at the moment 
why the Au sample follows the disorder theory, although it has $ql>1$ if calculated for transverse modes (Table 1). For longtitudinal modes it is 
still in the $ql <1$ limit.   
\section{Conclusions}
\sloppy
We have for the first time obtained clear evidence that the electron-phonon scattering rate scales with temperature as $1/\tau _{\rm e-p}\propto 
T^{4}$ in disordered evaporated Cu and Au thin films. This power law corresponds to electrons scattering from phonons mediated by disorder vibrating 
together with the phonon mode. In contrast, e-p scattering in the presence of static disorder leads to $1/\tau _{\rm e-p}\propto T^2$, a result that 
has been confirmed in several materials and samples \cite{Lin}. 

We stress that the theoretical result $1/\tau _{e-p}\propto T^{4}$ is valid for a simple metal with $ql<1$ and little nonvibrating disorder, and with 
3D phonons coupled to electrons by the deformation potential. We believe most earlier experiments do not satisfy all these conditions. 

\begin{acknowledgement}
  We thank D.-V. Anghel, A. N. Cleland, J. P. Pekola, H. Pothier, A. Savin and A. Sergeev for discussions.  This work was supported by the Academy of 
Finland under the Finnish Center of Excellence Program 2000-2005 (Project No. 44875), and by the Academy of Finland project No. 105258.    

\end{acknowledgement}

\end{document}